\documentclass[twocolumn,notitlepage,prl,superscriptaddress,showpacs]{revtex4-1}
\usepackage{hyperref}
\usepackage[usenames,dvipsnames]{color}
\usepackage{amsmath}
\usepackage[english]{babel}
\usepackage{amssymb}
\usepackage{graphicx}
\usepackage{epstopdf}
\usepackage{tabularx}
\usepackage[]{todonotes}

\begin{document}
\title{Superfluid ground state phase diagram of the two-dimensional Hubbard model in the emergent Bardeen-Cooper-Schrieffer regime}

\author{Fedor \v{S}imkovic IV}
\email{fsimkovic@gmail.com}
\affiliation{Department of Physics, King’s College London, Strand, London WC2R 2LS, United Kingdom}
\affiliation{Centre de Physique Th\'eorique, \'Ecole Polytechnique, CNRS, Universit\'e Paris-Saclay, 91128 Palaiseau, France}
\affiliation{Coll\`{e}ge de France, 11 place Marcelin Berthelot, 75005 Paris, France.}

\author{Youjin Deng}
\affiliation{Hefei National Laboratory for Physical Sciences at Microscale and Department of Modern Physics, University of Science and Technology of China, Hefei, Anhui 230026, China}
\affiliation{MinJiang Collaborative Center for Theoretical Physics, Department of Physics and Electronic Information Engineering, Minjiang University, Fuzhou, Fujian 350108, China}

\author{Evgeny Kozik}
\email{evgeny.kozik@kcl.ac.uk}
\affiliation{Department of Physics, King’s College London, Strand, London WC2R 2LS, United Kingdom}

\date{\today}

\begin{abstract}
In nonperturbative regimes, the superfluid instability in the two-dimensional Hubbard model can be described by an emergent BCS theory with small effective pairing constants.
We compute the effective couplings using a controlled bold-line diagrammatic Monte Carlo approach, which stochastically sums all skeleton Feynman diagrams dressed in the one- and two-particle channels to high expansion orders, and map out the resulting superfluid ground-state phase diagram in a range of next-nearest-neighbor hopping $0 \leq t^{\prime} \leq 0.3t$, interaction strength $0 \leq U \leq 3t$, and lattice filling $0 \leq n \leq 2$. The phase diagram is dramatically transformed in the hole-doped region and becomes particularly rich at larger doping and $t'$.
At $t'=0.3$, the weak-coupling picture with the dominant triplet pairing sharply peaked at $n \approx 0.73$ due to the Van Hove singularity is replaced by a plateau of the singlet $d_{x^2-y^2}$ paring, while for $U \gtrsim 3t$ the effective couplings are consistent with the $d_{x^2-y^2}$ high-temperature superconductivity in the hole-doped region near cuprates' optimal doping.
\end{abstract}


\maketitle

The fermionic Hubbard model \cite{hubbard1963electron,anderson1963theory,anderson1997theory,lee06hightc} is the simplest microscopic model of interacting electrons in solids:
\begin{equation}
H = \sum_{k,\sigma} \left(\epsilon_{k} -\mu\right)c_{k\sigma}^\dagger c_{k\sigma}+U\sum_i n_{i\uparrow}n_{i\downarrow}.
\label{H}
\end{equation}
Here $\mu$ is the chemical potential, $k$ in the momentum, $U$ is the on-site repulsion strength, $i$ labels the lattice sites, and the (square lattice) dispersion is given by
$$
\epsilon_k=-2t\left[\cos(k_x)+\cos(k_y)\right]-4t^{\prime}\cos(k_x)\cos(k_y),
$$
where $t$ and $t^{\prime}$ are the nearest- and next-nearest-neighbor hopping amplitudes ($t=1$ in our units), respectively. It is a workhorse of condensed matter theory, used for understanding a plethora of macroscopic quantum phenomena, such as the metal-to-insulator transition~\cite{georges1996dmft}, ferromagnetism and antiferromagnetism~\cite{hubbard1963electron,anderson1963theory}, and high-temperature superconductivity~\cite{anderson1997theory}. It is also the main testbed for novel computational approaches to correlated lattice electrons~\cite{leblanc2015solutions}, and a rare example of a paradigmatic model of many-body physics amenable to precise experimental realization, in particular with ultracold atoms in optical lattices \cite{Bloch:2005uv,Lewenstein:2007hr,Greif:2015bg,Cheuk:2016kq,Mazurenko:2017ec,Brown:2017dy, Nichols:2019iq}.

The Hubbard model on the square lattice and its relation to layered copper-oxide materials (cuprates) have been subject to particular scrutiny, with the grand goal of shedding light on high-temperature superconductivity~\cite{anderson1997theory}. However, a close competition between a multitude of superfluid and magnetic orders makes it a major, largely still unsolved, problem.

A numerically exact phase diagram has been established (semi)analytically at vanishingly small interactions and/or low fillings~\cite{baranov1992superconductivity, chubukov1992pairing, chubukov1993kohn, hlubina1999phase, raghu2010superconductivity, romer2015pairing, simkovic2016ground, Kreisel2017weak_coupling}. In this picture, described by the generalized BCS theory, the pairing instability develops in a Fermi liquid due to a small Cooper-channel attraction
resulting from nontrivial momentum dependence of the scattering matrix on the Fermi surface. The ground state near half-filling (average density per site $n=1$) for all relevant $t^{\prime}$ ($0 \leq t^{\prime} \leq 0.5$) was found to be, similarly to cuprates, a $d_{x^2-y^2}$-wave superfluid, while the phase diagram at larger dopings becomes remarkably rich with $d_{xy}$-, $p$-, $g$-, and $s$-wave superfluids also realized at different densities~\cite{simkovic2016ground}.

The regime of weak effective attraction in the Cooper channel
extends to substantial values of $U$, where the calculation of the scattering matrix becomes essentially nonperturbative. It is the so-called \textit{emergent} BCS regime, which has recently become tractable with controlled accuracy directly in the thermodynamic limit by means of diagrammatic Monte Carlo techniques~\cite{deng2015emergent}. Without next-nearest-neighbor hopping ($t^{\prime}=0$), the emergent BCS regime has been found to extend at least up to $U \lesssim 4$ and $n \lesssim 0.8$, and a phase diagram has been obtained~\cite{deng2015emergent}, which differs qualitatively from the weak-coupling limit~\cite{hlubina1999phase, raghu2010superconductivity, simkovic2016ground}, while the $d_{x^2-y^2}$-pairing at larger densities is robust. Evidence of $d_{x^2-y^2}$-wave superconductivity with a particularly high $T_c$ has been provided at intermediate to strong coupling ($U \gtrsim 3$) by means of embedded quantum cluster methods~\cite{Maier2000superconductivity, Maier2005superconductivity, Lichtenstein2000superconductivity, Capone2006superconductivity, Aichhorn2006superconductivity, Kancharla2008superconductivity, Sordi12superconductivity, Zheng2016ground, Chen2015superconducting, romer2019pairing} as well as the functional renormalization group approach~\cite{Metzner2012fRG_review, Friederich2011fRG_superconductivity, Yamase2016fRG_superconductivity}. Yet, recent studies by advanced tensor network and quantum Monte Carlo methods have shown that, at least for $U \gtrsim 6$ and $n \gtrsim 0.8$, high-$T_c$ superconductivity is completely wiped out by inhomogeneous magnetic (stripe) phases~\cite{Zheng2017stripe, Ido2018no_superconductivity, Qin2019absence_of_superconductivity}. Thus, the question of whether the two-dimensional Hubbard model with $t^{\prime}=0$ supports high-$T_c$ superconductivity at all remains open.

On the other hand, a minimal tight-binding model of high-$T_c$ cuprates must feature a significant next-nearest-neighbor hopping $0.1 \lesssim t^{\prime}\lesssim 0.3$~\cite{Pavarini01, tanaka2004effects, markiewicz2005one}, which is known to essentially alter the phase diagram already in the weak-coupling limit~\cite{hlubina1999phase, raghu2010superconductivity, simkovic2016ground}. Until now, however, a reliable phase diagram for nonzero values of $t^{\prime}$ and substantial couplings has remained unknown.
Moreover, at a nonzero $t^{\prime}$, the Fermi surface for certain fillings $n=n_{\text{VH}}$ features a Van Hove singularity in the density of states without being fully nested. Such fillings could favor Cooper pairing and substantially enhance the corresponding $T_c$, potentially explaining its peak at optimal doping \cite{dessau1993key, abrikosov1993experimentally, gofron1993k, gofron1994observation, king1994observation, Markiewicz_VH_review}, but the effects of $n_{\text{VH}}$ at nonperturbative interactions have not yet been reliably understood.


\begin{figure}[!htb]
\centering
\includegraphics[width=1.0\linewidth]{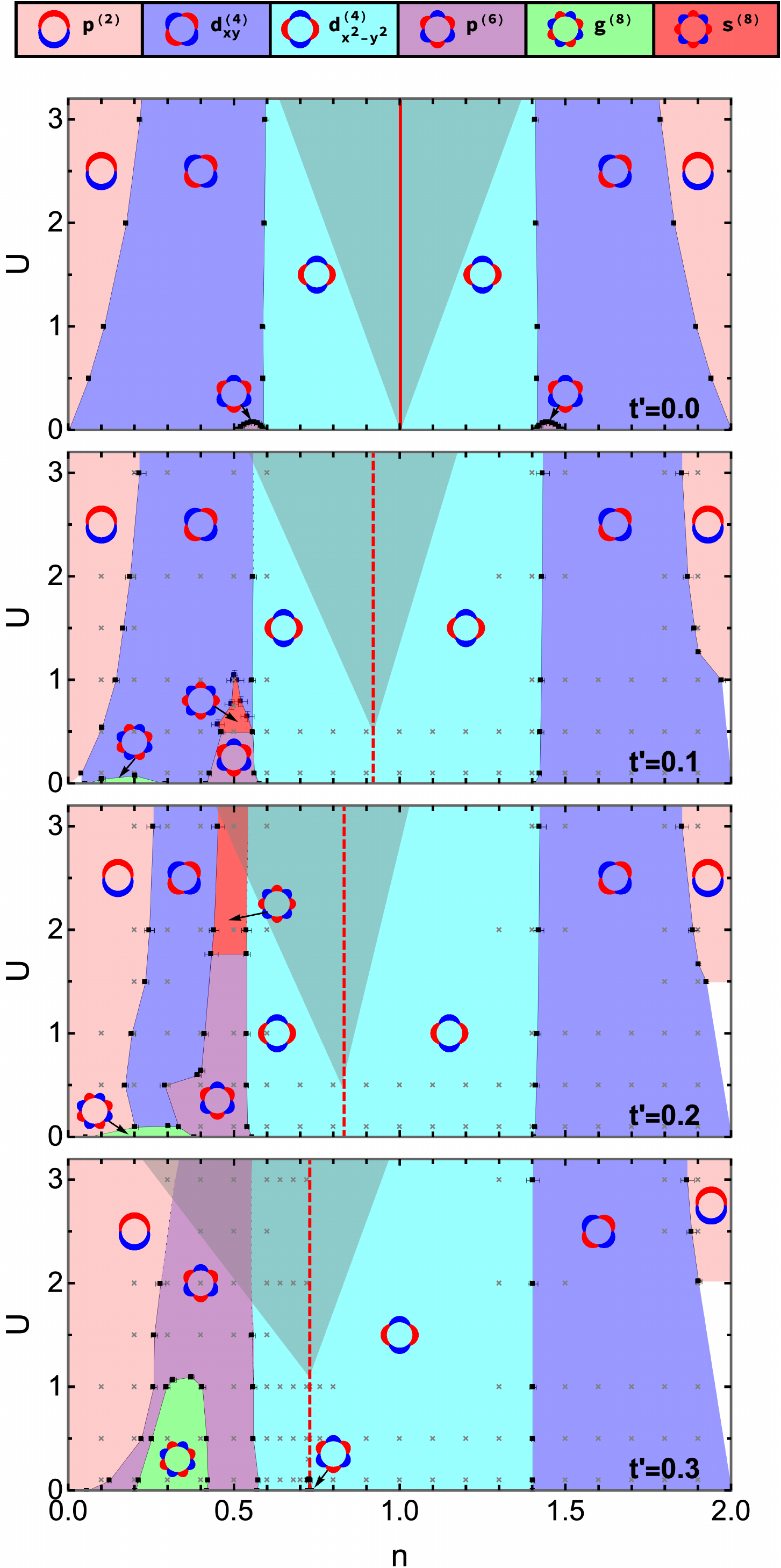}
\caption{Superfluid ground-state phase diagram for multiple values of next-nearest-neighbor hopping $t^{\prime}$. The $t^{\prime}=0$ results are adapted from Ref.~\cite{deng2015emergent}. In the gray-shaded regions the leading pairing constants become large ($\gtrsim 0.1$), potentially violating the condition $\lambda^*_S \ln(E_F/\xi_*) \ll 1$ and implying high $T_c$, while poor convergence with diagram order suggests competing instabilities. The red line is the antiferromagnetic ground state at $t^{\prime}=0$, dashed lines mark $n_{\text{VH}}$. Gray crosses are the points for which the calculations were performed.
}
\label{phase_diagram_tp}
\end{figure}

Here we study the emergent BCS regime of the Hubbard model on the square lattice in a range of next-nearest-neighbor hopping $0 \leq t^{\prime} \leq 0.3$, interaction strength $0 \leq U \leq 3$, and lattice filling $0 \leq n \leq 2$, and obtain the effective BCS couplings with controlled accuracy by the bold diagrammatic Monte Carlo (BDMC) technique~\cite{van2012feynman, deng2015emergent, VanHoucke2019BDMC}. In our approach, introduced in Ref.~\cite{deng2015emergent}, all Feynman diagrams for the irreducible in the Cooper channel vertex in terms of the self-consistently determined fully dressed one- and two-particle propagators are summed numerically exactly up to a sufficiently high expansion order (defined as the number of two-particle propagators in the diagram) until convergence, while the irreducible vertex determines the effective pairing constants $\lambda_S$ in each symmetry sector $S$ ($S=s, g, d_{xy}, d_{x^2-y^2}, p_x, p_y$ on the square lattice). The resulting superfluid phase diagram, Fig.~\ref{phase_diagram_tp}, which is also unbiased under an additional natural assumption (to be discussed below), transforms dramatically with $t^{\prime}$: the $p$-wave
regions with two different nodal structures expel the $d_{xy}$ phase on the hole-doped side, while new $s$ and $g$ phases appear and the region around $n_{\text{VH}}(t^{\prime})$ with the highest pairing strengths remains occupied by the $d_{x^2-y^2}$-wave superfluid which is prevails for densities $n>0.55$ regardless of the values of $t'$ and $U$. At $t^{\prime}=0.3$ most relevant for cuprates, we find that the weak-coupling scenario~\cite{raghu2010superconductivity, simkovic2016ground}---in which the maximum of $T_c$ with doping is due to the $p$-wave pairing that is sharply peaked at $n=n_{\text{VH}}\approx 0.73$---is transformed already at $U \sim 0.5$: the $d_{x^2-y^2}$ pairing becomes dominant and develops a plateau at $n_{\text{VH}}$ that stretches to the vicinity of cuprates' optimal doping $n \sim 0.8$. A self-consistent diagrammatic expansion up to orders as high as 7 is necessary to capture with controlled accuracy the correct behavior already at $U\sim 1$, demonstrating that the emergent BCS regime of the 2D Hubbard model is not amenable to perturbative treatment~\footnote{A single bold-line diagram is already nonperturbative containing an infinite series in the powers of $U$.}. The largest (among all symmetry sectors) pairing constant rises steeply with density, reaching $\approx 0.25$ at $n = n_{\text{VH}}$ and $U=3$, where it corresponds to the $d_{x^2-y^2}$ pairing, and marking the regime of strong Cooper-channel coupling, consistent with high-temperature superconductivity. At larger densities, $\lambda_{d_{x^2-y^2}}$ increases with diagram order without signs of saturation, implying a competition between the superfluid and magnetic instabilities.

 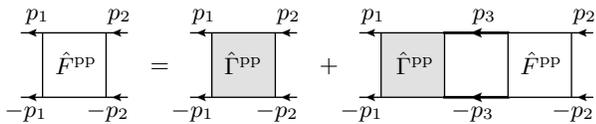
\begin{figure}[!ht]
  \centering
  \begin{picture}(50,30)
  \put(-85,0){\includegraphics[height=1cm]{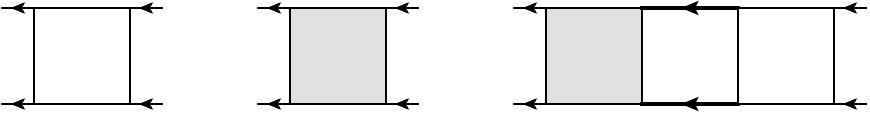}}
\put(-60,-5){$-p_2$}\put(-91,-5){$-p_1$}\put(-52,32){$p_2$}\put(-83,32){$p_1$}
\put(-72,11){$\hat{F}^{\text{pp}}$}
\put(-36,12){$=$}
\put(4,-5){$-p_2$}\put(-28,-5){$-p_1$}\put(12,32){$p_2$}\put(-20,32){$p_1$}
\put(-8,11){$\hat{\Gamma}^{\text{pp}}$}
\put(28,12){$+$}
\put(116,-5){$-p_2$}\put(36,-5){$-p_1$}\put(78,-5){$-p_3$}\put(124,32){$p_2$}\put(44,32){$p_1$}\put(86,32){$p_3$}
\put(57,11){$\hat{\Gamma}^{\text{pp}}$}
\put(104,11){$\hat{F}^{\text{pp}}$}
\end{picture}
\caption{The Bethe-Salpeter equation. Here $p_i \equiv \left( \xi_i, \mathbf{k}_i \right)$ and summation over $\xi_3$ and integration over $\mathbf{k}_3$ is assumed.}
\label{bse}
\end{figure}

BDMC enables numerically exact calculation of the irreducible in the particle-particle (Cooper) channel vertex $\Gamma^{\text{pp}}$ directly as a sum of all possible four-point diagrams that cannot be split into disconnected pieces by cutting two particle lines~\citep{AGD}. $\Gamma^{\text{pp}}$ drives the superfluid instability, which is marked by divergence at $T_c$ of the full four-point vertex $F^{\text{pp}}$ according to the Bethe-Salpeter equation (BSE), shown diagrammatically in Fig.~\ref{bse}. Here the solid lines are the many-body Green's functions $G[p=(\xi, \mathbf{k})]$ in terms of the Matsubara frequencies $\xi$ and momenta $\mathbf{k}$. A direct solution of the BSE is not feasible because of its vast data content and essential nonlinearity. Nonetheless, when $\Gamma^{\text{pp}}$ is small, it becomes tractable semi-analytically due to the separation between the Fermi energy $E_F$, the typical frequency scale $\xi_*$ at which $\Gamma^{\text{pp}}$ varies and temperature: $T \ll \xi_* \ll E_F$. It comes from the requirement that, for small $\Gamma^{\text{pp}}$, the divergence of $F^{\text{pp}}$ in the BSE must come from a large contribution of $G(p_3)G(-p_3)$, which, being summed over frequencies, grows at best logarithmically slowly with $E_F/T$, provided $G(p)$ is that of a fully developed Fermi liquid. In this case, with logarithmic accuracy, at $T \sim T_c$, the BSE reduces to (see, e.g., Ref.~\cite{deng2015emergent})
\begin{equation}
F^{\text{pp}}_{\hat{k}_1,\hat{k}_2} \approx \Gamma^{\text{pp}}_{\hat{k}_1,\hat{k}_2}+ \ln\frac{c E_F}{T} \int \Lambda_{\hat{k}_1,\hat{k}_3} F^{\text{pp}}_{\hat{k}_3,\hat{k}_2} d \hat{k}_3,
\label{BSE_FL}
\end{equation}
where all the functions are taken at vanishing frequencies and projected onto the Fermi surface, e.g., $F^{\text{pp}}_{\hat{k}_1,\hat{k}_2} \equiv F^{\text{pp}} (\mathbf{k}_1 = \mathbf{k}_F (\hat{k}_1), \xi_1 \rightarrow 0;\mathbf{k}_2 = \mathbf{k}_F (\hat{k}_2), \xi_2 \rightarrow 0)$ (with $\mathbf{k}_F$ the Fermi momentum, $\hat{k}=\mathbf{k}/|\mathbf{k}|$), $c$ is a constant of order unity, and the matrix $\Lambda_{\hat{k}_1,\hat{k}_1}$ is straightforwardly related to $\Gamma^{\text{pp}}_{\hat{k}_1,\hat{k}_2}$ via the Fermi surface parameters (see Ref.~\cite{deng2015emergent} for details). Thus, (block-)diagonalizing the matrix $\Lambda_{\hat{k}_1,\hat{k}_1}$ in the basis of the irreducible representations of the point group of the lattice (see, e.g., Ref.~\cite{simkovic2016ground}), one finds that $F^{\text{pp}}$ diverges at $T_c = c E_F e^{-1/\lambda_S}$, where $\lambda_S$ is the largest positive (attractive in these notations) eigenvalue of $\Lambda$. The eigenvector corresponding to $\lambda_S$ determines the spatial structure of the order parameter just below the superfluid transition.


Note that Eq.~(\ref{BSE_FL}) reduces the dependence of the BSE on the full $\Gamma^{\text{pp}}(p_1,p_2)$ to its zero-frequency part and only on the Fermi surface. This is a major simplification for practical calculations, since only constants $\lambda_S$ need to be computed by BDMC, which allows us to achieve small statistical error bars. Systematic relative corrections to the phase diagram lines determined in this way are on the order of $\lambda^*_S \ln(E_F/\xi_*)$~\cite{Chubukov2019_implicit_renormalization}, where $\lambda^*_S$ is the typical variation with frequencies of the full matrix $\Lambda(\xi_1, \xi_2, \mathbf{k}_1, \mathbf{k}_2)$ projected onto the sector $S$. Since computing the full frequency dependence of $\Lambda$ in practice requires unrealistic computational resources, here we can only estimate the corresponding systematic error: In all $S$ except the nodeless $s$ wave (for which the high-frequency effective coupling reduces to the bare $U$), $\lambda^*_S \sim |\lambda_S|$ since the effective coupling vanishes at high frequencies. Given that the obtained values of $\lambda_S$ for the claimed superfluid phases do not exceed a few percent, in view of $T_c \ll \xi_* \ll E_F$ it is reasonable to assume that $\lambda^*_S \ln(E_F/\xi_*) \ll 1$. Under this assumption, the phase diagram in Fig.~\ref{phase_diagram_tp} is unbiased, except the gray-shaded regions near $n_{\text{VH}}$ discussed separately.

\begin{figure*}[!ht]
\centering
\vspace{-0.0cm}
\includegraphics[width=0.95 \textwidth]{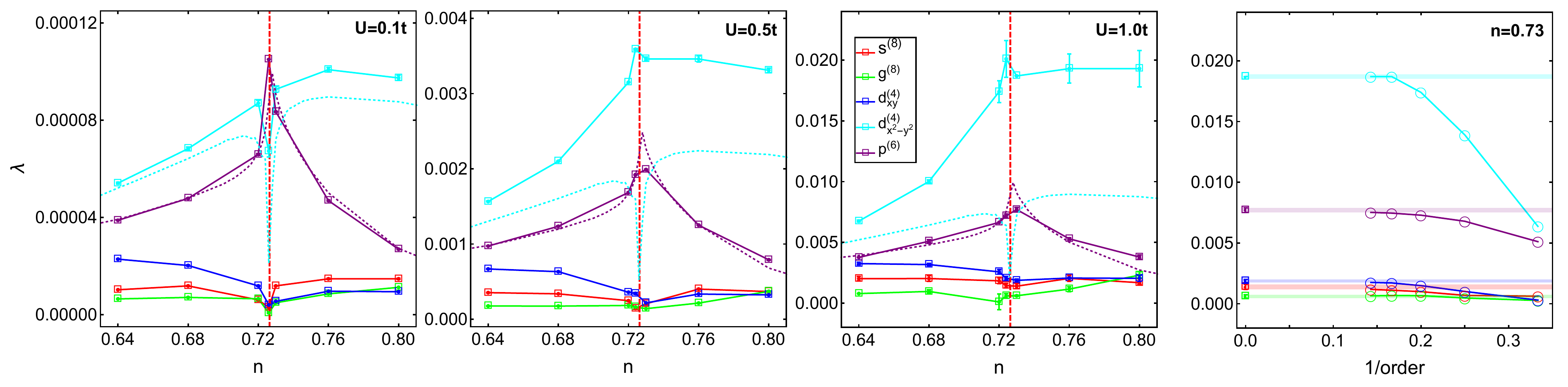}
\vspace{-0.0cm}
\caption{Left: Effective coupling strengths corresponding to different superfluid states for density values near $n_{\text{VH}}(t^{\prime}=0.3)=0.726$ (dashed, red) and interaction strengths $U=\{0.1,0.5,1.0\}$. Dotted lines represent second order bare perturbation theory results. Right: Partial sums for effective coupling strengths as a function of inverse diagram order, for $n=0.73$, $t^{\prime}=0.3$ and $U=1.0$. Horizontal lines represent results extrapolated to infinite expansion order.}
\label{van_hove_finite_U}
\end{figure*}

In order to obtain numerically exact values for $\lambda_S$ with acceptable error bars we must ensure that the computed diagrammatic series can be reliably extrapolated to infinite order within the diagram orders accessible by BDMC in reasonable time. Because at the two-particle level the expansion is renormalized only in the particle-particle channel, while reducible diagrams in other channels are summed explicitly, competing instabilities in other channels, such as, e.g., towards antiferromagnetic or stripe phases, manifest themselves as lack of convergence. Lack of convergence can also indicate~\cite{Rossi2016_shifted_action} proximity to the branching point of the Luttinger-Ward functional~\cite{kozik2015nonexistence}, beyond which the skeleton series is not reliable. We were able to evaluate the series for $\Gamma^{\text{pp}}_{\hat{k}_1,\hat{k}_2}$ with the fully self-consistent determination of the one- and two-particle propagators up to order 7 with under a million CPU hours available to us. We then obtained boundaries between different states in Fig.~\ref{phase_diagram_tp} from linearly interpolating $\lambda_S$ on a mesh of calculated points.


Figure~\ref{phase_diagram_tp} shows the resulting ground-state phase diagram for $t^{\prime}= \{0.1, 0.2, 0.3\}$ (three lower panels) along with the $t^{\prime}= 0$ data from Ref.~\cite{deng2015emergent} (upper panel) for comparison. We denote the phases $S^{(n)}$, where $n$ is the number of nodes in the order parameter of the symmetry $S$, as exemplified in the top row in Fig.~\ref{phase_diagram_tp}. The singlet $d_{x^2-y^2}^{(4)}$ superfluid near half-filling is a distinctive feature of the 2D Hubbard model found by most calculations at larger $U$ as well~\cite{Maier2000superconductivity, Maier2005superconductivity, Lichtenstein2000superconductivity, Capone2006superconductivity, Aichhorn2006superconductivity, Kancharla2008superconductivity, Sordi12superconductivity, Zheng2016ground, Chen2015superconducting} in relation to cuprates. On the electron-doped side ($n>1$), changes with increasing $t^{\prime}$ are minimal: the tiny $p^{(6)}$ region disappears already at $t^{\prime}=0.1$, as predicted from weak coupling~\cite{simkovic2016ground}, the $p^{(2)}$ region near $n=2$ shrinks noticeably towards higher $U$ in favor of $d_{x^2-y^2}^{(4)}$, while the boundary between $d_{xy}^{(4)}$ and $d_{x^2-y^2}^{(4)}$ is curiously insensitive to $t^{\prime}$ or $U$. The $p^{(2)}$-$d_{xy}^{(4)}$ boundary for $U \to 0$, $n \to 2$ can be obtained analytically, as in the case of $t^{\prime}=0$~\cite{chubukov1992pairing, chubukov1993kohn}, which is left for future work.

On the hole-doped side ($n<1$), the boundary of the $d_{x^2-y^2}^{(4)}$ phase shifts only slowly to lower dopings, but the neighboring $d_{xy}$ is gradually replaced by $p^{(6)}$, which grows along the boundary from a tiny small-$U$ region at $t^{\prime}=0$, eventually expelling the $d_{xy}$ phase for all accessible $U$. New $s^{(8)}$ and $g^{(8)}$ phases appear at $t^{\prime}>0$. Both have eight nodes in the gap function but similarly to $d_{xy}^{(4)}$ and $d_{x^2-y^2}^{(4)}$ belong to different irreducible representations of the $D_4$ symmetry group and can be obtained from one another by a $\pi/4$ rotation. An $s^{(8)}$ region appears at $t^{\prime}>0$ wedging between $p^{(6)}$ and $d_{xy}$ at larger $U$ and its extent in the vertical ($U$-) direction grows very rapidly with increasing $t^{\prime}$. 
Interestingly, the $s^{(8)}$ phase appears only at essentially finite $U$ and, as such, is entirely nonperturbative. A region of $g^{(8)}$ grows with $t^{\prime}$ from weak coupling at small densities, remaining limited to $U \lesssim 1$. The $p^{(2)}$-wave phase spreads to higher densities with increasing $t^{\prime}$ and $U$, which together with the growing $p^{(6)}$ squeezes $d_{xy}$ out, so that by $t^{\prime}=0.3$ the triplet superfluid dominates the larger doping ($n<0.6$) region of the diagram and $d_{xy}$ disappears.


\begin{figure}[!htb]
\centering
\includegraphics[width=1.0\linewidth]{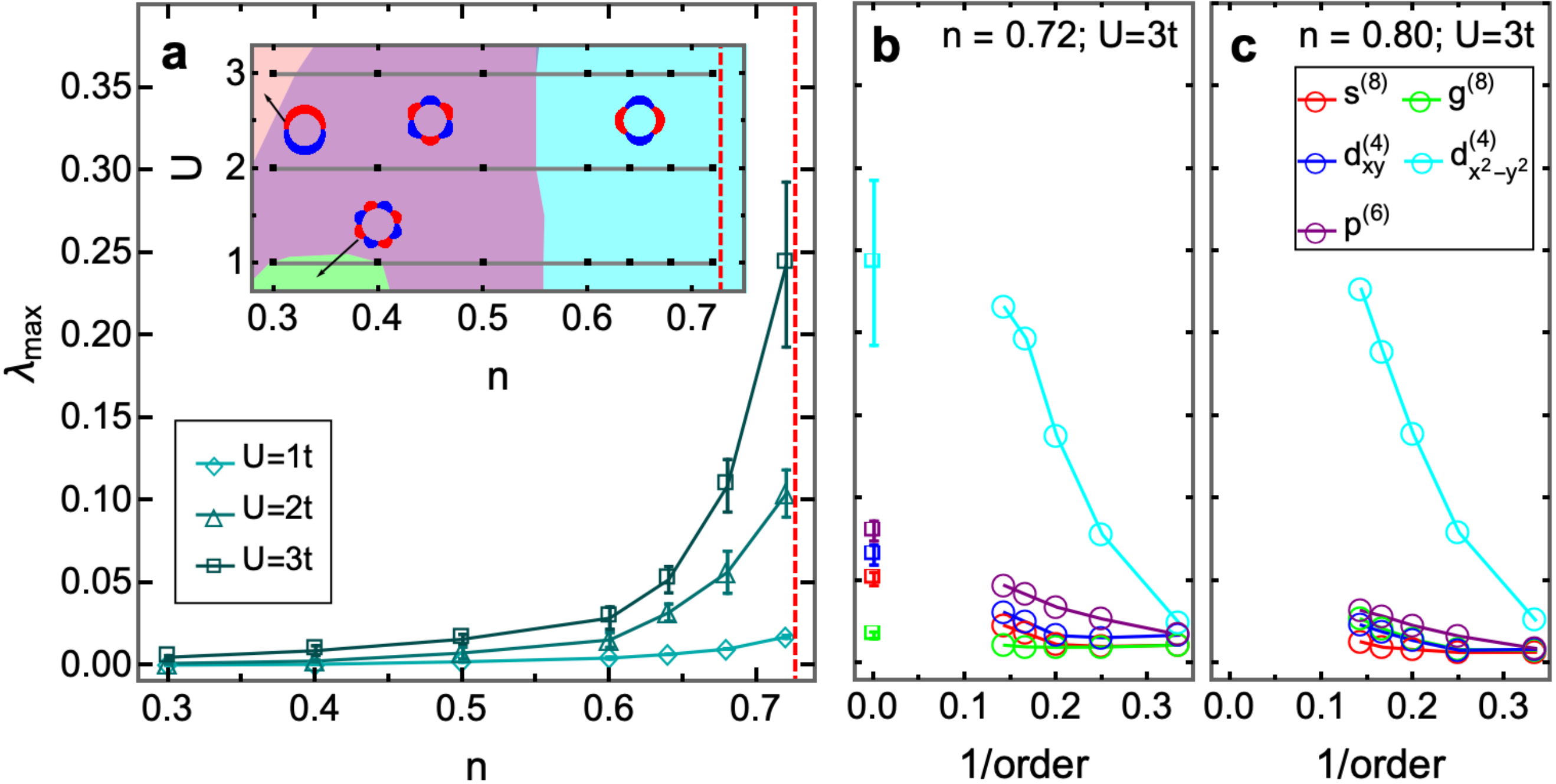}
\caption{ (a) Rapid increase of the maximal effective coupling $\lambda_\mathrm{max}$ on approach to $n_{\text{VH}}$ at $t^{\prime}=0.3$; inset: the gap function type for each point $(n, U)$ in the notations of Fig.~\ref{phase_diagram_tp}. (b) Nature of the series convergence at $n=0.72$, $U=3$ for the effective coupling in the leading symmetry sectors, and the corresponding values obtained by a linear extrapolation from the results with their error bars of the two highest expansion orders with a conservative estimate of the systematic error. (c) Slow convergence or divergence of the series at $U=3$ near cuprates' optimal doping.}
\label{lambda_max}
\end{figure}

Near $n_{\text{VH}}=n(\mu=4t^{\prime}) \approx 0.726$, convergence of the diagrammatic series for $\Gamma^{\text{pp}}$ up to order 7 becomes slow at $U > 1$, with the effective couplings growing substantially with diagram order. This behavior is consistent with the general picture of the effects of the singular density of states on the Fermi surface suggested by mean-field~\cite{hirsch1985two, lin1987two, irkhin2001effects, hankevych2003a} and renormalization group studies~\cite{alvarez1998superconducting, halboth2000, honerkamp2001a, katanin2003, igoshev2010incommensurate}, in which $d_{x^2-y^2}$ pairing is fuelled by the competition with commensurate or incommensurate magnetic (stripe) phases that win over at a larger $U$~\cite{Zheng2017stripe, Ido2018no_superconductivity, Qin2019absence_of_superconductivity}. This is the regime where the pairing constants are typically the largest, potentially explaining the maximum of $T_c$ in cuprates \cite{Markiewicz_VH_review}. We study it specifically at $t^{\prime}=0.3$ relevant to cuprates.
Figure~\ref{van_hove_finite_U} shows $\lambda_S$ for $U=\{0.1, 0.5, 1.0\}$ as well as the typical dependence on diagram order. At $U=0.1$ the scenario is qualitatively similar to weak coupling \cite{raghu2010superconductivity,simkovic2016ground}: the $d_{x^2-y^2}^{(4)}$ pairing experiences a dip around $n_{\text{VH}}$ while the $p^{(6)}$ one is sharply peaked making it the leading instability. However, already at $U=0.5$, the leading state around $n_{\text{VH}}$ (at least within our resolution $\delta n = 0.001$) is always $d_{x^2-y^2}^{(4)}$, the $p^{(6)}$ pairing being still peaked but twice as low. The couplings increase with $U$, but the qualitative shape of $\lambda_{d_{x^2-y^2}}(n)$ is robust:  it grows with density and plateaus for $n > n_{\text{VH}}$ until at least $n \sim 0.8$.


Of special interest is the regime of $U>1$, where $\{\lambda_S\}$ grow rapidly and become very large on approach to $n_{\text{VH}}$ (gray-shaded regions in Fig.~\ref{phase_diagram_tp}). Figure~\ref{lambda_max}(a) reveals that the maximal among all symmetry sectors effective coupling $\lambda_\mathrm{max}$ reaches $\approx 0.25$ at $U=3$ and $n=n_{\text{VH}}$, where it is of the $d_{x^2-y^2}$ character. Remarkably, it grows steeply with the diagram order [Fig.~\ref{lambda_max}(b)], so that a low-order perturbation theory would underestimate $\lambda_\mathrm{max}$ by an order of magnitude, wrongly concluding that $U=3$ is a regime of weak correlations with an exponentially small $T_c$. In contrast, the (emergent) BCS estimate of $T_c$ for the actual $\lambda_\mathrm{max}\approx 0.25$ in a real system is $\sim$~200K, consistent with $d_{x^2-y^2}$ high-$T_c$ superconductivity. At larger densities, e.g., $n=0.8$ [Fig.~\ref{lambda_max}(c)], $\lambda_\mathrm{max}$ does not show saturation within the accessible diagram orders, which only allows us to establish its lower bound $\lambda_\mathrm{max} > 0.25$. The lack of convergence is consistent with a competing instability towards a magnetic (antiferromagnetic or stripe) phase, which is known to dominate over superconductivity at these densities at least for $t^\prime=0$, $U \gtrsim 6$~\cite{Zheng2017stripe, Ido2018no_superconductivity, Qin2019absence_of_superconductivity}.

Our results thus suggest that $d_{x^2-y^2}$ high-$T_c$ superconductivity driven by strong magnetic fluctuations could be realized at $t'=0.3$ already at $U$ as low as $\sim 3$, and $n \gtrsim n_{\text{VH}}$ near cuprates' optimal doping. The large values of the effective couplings almost certainly violate $\lambda^*_S \ln(E_F/\xi_*) \ll 1$, meaning that the emergent BCS theory that relies solely on $\lambda_S$, as well as the Fermi-liquid character of the normal state, becomes questionable. The recently proposed implicit renormalization approach~\cite{Chubukov2019_implicit_renormalization} addresses this problem, which, combined with accurate description of correlations near magnetic instabilities~\cite{simkovic2017magnetic, Simkovic2018crossover, aaram:spin_charge2019}, could enable controlled studies of superconductivity in this regime in the future.

\acknowledgements{We are grateful to A.~Chubukov, B.~Svistunov, and N.~Prokofiev for discussions of the results and X.-W.~Liu for his help with data aggregation. This work was supported by the EPSRC through Grant No. EP/P003052/1 and the Simons Foundation as a part of the Simons Collaboration on the Many-Electron Problem. YD acknowledges the support from National Natural Science Foundation of China (Grant No.11625522) and the Ministry of Science and Technology of China (Grant No. 2018YFA0306501).}

\bibliographystyle{apsrev4-1}
\bibliography{refs.bib}

\end{document}